\documentclass[11pt,a4paper,reqno]{amsart}

\usepackage{amsmath}
\usepackage{amsfonts}
\usepackage{amssymb}
\usepackage{bm}
\usepackage{graphicx}
\usepackage{amsbsy}
\usepackage{subfig}
\usepackage{dsfont}
\usepackage{comment}

\newtheorem{theorem}{Theorem}

\newtheorem{remark}[theorem]{\it \bf{Remark}\/}

\newcommand{\CC}{\mathbb{C}}
\newcommand{\RR}{\mathbb{R}}

\newcommand{\Ss}{\mathbb{S}}

\newcommand{\eps}{\varepsilon}

\newcommand{\hH}{{\mathcal H}}

\newcommand{\ual}{\boldsymbol\alpha}

\newcommand{\um}{\boldsymbol\mu}

\def\RR{{\mathrm{ I~\hspace{-1.15ex}R}}}

\let\ds\displaystyle
\def\be{\begin{equation}}
\def\ee{\end{equation}}
\def\qquad{{\quad\quad}}

\title[A quantum particle in a quantum environment]{A model of a quantum particle in a quantum environment: a numerical study}

\author[R. Carlone, R. Figari, C. Negulescu]{Raffaele Carlone, Rodolfo Figari, Claudia Negulescu}

\address{Raffaele Carlone, Universit\`a Federico II di Napoli, Dipartimento di Matematica e Applicazioni "R. Caccioppoli", MSA I-80126 Napoli, Italy.}
\email{raffaele.carlone@unina.it}
\address{Rodolfo Figari, Universit\`a Federico II di Napoli, Dipartimento di Fisica e INFN Sezione di Napoli, MSA I-80126 Napoli, Italy.}
\email{figari@na.infn.it}
\address{Claudia Negulescu, Universit\'e de Toulouse \& CNRS, UPS, Institut de Math\'ematiques de Toulouse UMR 5219, F-31062 Toulouse, France.}
\email{claudia.negulescu@math.univ-toulouse.fr}

\usepackage{color}

\begin{document}

\maketitle

\begin{abstract}
We define and investigate, via numerical analysis, a one dimensional toy-model of a cloud chamber. An energetic quantum particle, whose initial state is a superposition of two identical wave packets with opposite average momentum, interacts during its evolution and exchanges (small amounts of) energy with an array of localized spins. Triggered by the interaction with the environment, the initial superposition state turns into an incoherent sum of two states describing the following situation: or the particle is going to the left and a large number of spins on the left side changed their states, or the same is happening on the right side. This evolution is reminiscent of what happens in a cloud chamber where a quantum particle, emitted as a spherical wave by a radioactive source, marks its passage inside a supersaturated vapour-chamber in the form of a sequence of small liquid bubbles arranging themselves around a possible  classical trajectory of the particle.
\end{abstract}

\bigskip


\section{Introduction} \label{sec:intro}
In this paper we investigate numerically the dynamics of a quantum particle interacting with a quantum environment. More precisely, we consider the semi-classical limit regime of the dynamics.  With this we mean that the average initial kinetic energy of the particle is assumed to be very large with respect to the energy exchanged by the particle with the environment. \\

The paradigmatic physical system we have in mind is the Wilson cloud chamber, the prototype of a tracking chamber for elementary particle detection. Inside the chamber, a very energetic $\alpha$-particle, emitted in a radially symmetric way by a radioactive source, ionizes atoms of a super-saturated vapor. In turn, the ionized atoms become condensation nuclei, triggering the formation of sequences of liquid drops. The tracks one observes in real experiments look quite explicitly as classical particle trajectories. \\

In the early days of quantum mechanics Darwin, Heisenberg and Mott were the first to point out the seemingly paradoxical circumstance  of an initial radially symmetric quantum state evolving into wave packets concentrated around classical trajectories. In different ways, they suggested that the problem could be faced taking into consideration that the wave function does not evolve in real space, but rather in the configuration space of the entire quantum system. This somehow obvious but extremely far-reaching intuition, exploited by Mott in his seminal work (\cite{mo}), remained unnoticed for decades. \\

More recently, researchers analyzed the cloud chamber problem focusing on different aspects of the interaction of a microscopic quantum system with a macroscopic one.  
\begin{itemize}
\item[Decoherence :] the initial state of the $\alpha$-particle can be seen as a superposition of coherent states each one having a well localized momentum direction. The superposition is initially strongly coherent: in absence of any interaction with the environment, two coherent states might interfere in a double-slit experiment. On the other hand, coherent states heading in different directions generate macroscopic ionization in different regions of the environment. Due to this particle-environment interaction, the state of the whole system evolves into an incoherent superposition of states supported in distant regions of the environment configuration space, in such a way that any interference effect is prevented. For a general presentation of the decoherence phenomenon see e.g. (\cite{hs}) and references therein. For details on collisional decoherence in a tracking chamber we refer to the book (\cite{ft}).

\item[Non demolition measure :] a microscopic system (e.g. a quantum particle) is said to undergo a non demolition measure if there is a basis of its states which are left unchanged by the interaction with the probe (the measurement apparatus). At the same time, in order to work as a measurement apparatus, the probe should evolve in different final states for different particle states in the basis. Some authors analyzed recently the effect of repeated non demolition measures and its relation with the ``collapse" of the wave function in a quantum measurement (\cite{bb}). In this language, the process described above can be rephrased in the following way: each coherent state with a well defined momentum direction is an element of the basis. For a very high average initial energy (semi-classical conditions) each state of the basis evolves almost freely in each weakly-inelastic scattering process, whereas the environment reacts in different ways according to the average momentum direction of each coherent state. Repeated scattering processes will bring to the collapse of the wave function on one of the states in the superposition.
 \end{itemize}

\noindent Besides the fundamental aspects mentioned before, other motivations for understanding the dynamics of decoherence come from  several technological applications exploiting quantum coherence, such as quantum computers, electron spin resonance, nuclear magnetic resources and so on. Efforts are hence made in order to try to curb the destructive role of decoherence in such applications.\\

\noindent
To model and give rigorous mathematical results concerning the dynamics of a quantum particle in a many body quantum environment is a difficult task. In the second half of the last century, few attempts to define simplified solvable particle-environment models  were made, starting from the seminal papers (\cite{hep}) and (\cite{jz}). See also (\cite{gjkksz}).  
More recent investigations on the same line are (\cite{dft1}, \cite{dft2}, \cite{ccf2}, \cite{figt}, \cite{ft}, \cite{rt}, \cite{te}, \cite{ahn}). In all these models the environment is  made of either two or more level ``atom''-arrays (spins or oscillators) or of a gas of light particles.

\noindent
In the following, we present a simple one dimensional model of a cloud chamber. The environment consists of an array of two level quantum systems (sometimes referred to as atoms or spins) kept in fixed positions. The particle initial state is made of two identical wave packets concentrated in the origin (where a radioactive source is located) and moving away from the origin with opposite average momentum. All the atoms are initially in their ground state. The interaction particle-environment is modelled by multi-channel point potentials (\cite{ccf}), allowing  energy exchanges between the particle  and the atoms. \\

It is one of the simplest particle-environment model, permitting a reasonable numerical study of the decoherence phenomenon. Numerical simulations of the whole system evolution will be presented and confirm that after the interaction process has taken place,  the solution, apart for negligible terms, has the form of an incoherent sum of three states describing alternative histories of the environment: either no significative atom excitation has taken place or only atoms on one side of the origin are found in an excited state. Furthermore, the numerical results show that the higher the number of  environment constituents is (spins or atoms), the more effective is the environment induced decoherence. This is a new and important {achievement} of the present work. A different model has been investigated in \cite{ahn} to study  the decoherence effects induced by successive two-body heavy-light particle interactions. The model presented here is somehow more realistic, as it examines the real dynamics of the whole system, where simultaneous particle-environment interactions are allowed.\\

The paper is organized as follows. 
Section \ref{sec1} contains a  concise introduction of the particle-environment model presented in (\cite{ccf2}).  In section \ref{sec2} we give details  on the space-time discretization of the above model, used in the numerical computation of the wave-function solutions. The results of the numerical simulations are presented in section \ref{sec3}. Comments on these results and achievable future extensions of this work are presented in the final section.

\section{Multi-channel point interactions} \label{sec1}

In the present section we introduce a simple one-dimensional mathematical model for the gas inside a Wilson-chamber. We analyze the dynamics of the wave-function $\Psi(t,x)$ representing the whole quantum-mechanical system, made of an $\alpha$-particle and its environment, having the role of detecting the particle passage.  The interaction particle-environment is modelled  via multi-channel point interactions.

\subsection{The model} \label{SEC11}

We consider a quantum  particle moving on a line and interacting with an array of $N$ localized spins. In mathematical terms, the time evolution of this toy model is given by the Schr\"odinger equation

\be \label{SCH}
\left\{
\begin{array}{l}
{\mathbf i}\,\hbar\, \partial_t\Psi=H\, \Psi\\[2mm]
\Psi(0, \cdot)=\Psi_0\,,
\end{array}
\right.
\ee
where $\Psi$ is the wave-function describing the whole quantum system, $\hbar$ is the Planck constant and $H:D(H) \subset \hH \rightarrow \hH$ is the Hamiltonian whose domain and action will be defined later on.\\

The space of the system states is the Hilbert-space $\hH=L^{2}(\RR)\otimes \Ss_{N}$. Here, $\,\,\displaystyle\Ss_{N}=\underbrace{\CC^{2}\otimes\dots\otimes\CC^{2}}_{\text{N times}}$ is the configuration space of the environmental spins, whereas $L^{2}(\RR)$ is the configuration space of the $\alpha$-particle. The $N$ spins are assumed to be localized in the positions $\,Y=\left\{y_1,\cdots,y_N\right\} $ with $ y_{j}\in \RR$ for all $j \in \mathcal{J} :=\{1, \ldots,N\}$. \\

\noindent
 Let us denote by  $\displaystyle\hat\sigma^{(3)}=\left(\begin{matrix}1 & 0 \\0 & -1\end{matrix}\right)$ the third Pauli matrix. The $j$-th spin state space $\CC^{2}$ can be viewed as the complex linear span of the spin eigenstates $\chi_{\sigma_{j}}$ corresponding to the eigenvalues $\sigma_j=\pm 1\,\,$ of $\,\,\hat\sigma^{(3)}$, and representing the ``spin up'' and ``spin down'' states. 

\noindent
The state space $\Ss_{N}$ of the entire spin array will be the complex linear span of the basis vectors $\chi_{\underline\sigma}=\chi_{\sigma_{1}}\otimes\cdots\otimes\chi_{\sigma_{N}}$, where ${\underline\sigma}:= (\sigma_1, \ldots, \sigma_N) \in {\mathcal S}:=\{ \pm 1 \}^N$ denotes one of the $M:=2^N$ possible spin configurations.\\

\noindent
With this notation, the system consisting of the quantum particle and the $N$ spins is described by the wave-function 

\be
\Psi=\sum_{\underline\sigma \in {\mathcal S}}  \psi_{\underline\sigma}\otimes\chi_{\underline\sigma}\,, \qquad \Psi\in\mathcal{H}\,,
\ee
 where the sum runs over all possible spin configurations of $\underline\sigma \equiv (\sigma_{1},\cdots,\sigma_{N})$. In this decomposition each $\psi_{\underline\sigma} \in L^{2}(\RR)$ represents the wave function of the particle when the spin configuration is $\underline\sigma$.

Alternatively, in the following we will use for the state of our quantum system the vectorial notation 

$$
\Psi=(\psi_{{\underline \sigma}_1}, \cdots, \psi_{{\underline \sigma}_{M}})^{t}=(\psi_{\underline \sigma})_{{\underline \sigma} \in \mathcal S} \in (L^2(\RR))^{M}\,.
$$
Correspondingly the Hamiltonian in  \eqref{SCH} will take the form of a $M \times M$ matrix.\\

\noindent
The dynamics of the system is governed by the total Hamiltonian
\begin{equation}\label{H}
H=H_0+H_I\,,
\end{equation}
which is decomposed into two distinct parts, $H_0$ describing the free independent evolution of the particle  and of the spins, and $H_ I$ describing the particle-environment interaction.  Furthermore, the interaction Hamiltonian $H_ I$ will be decomposed into two parts: a zero range interaction $H_D$, supported by the set of the spin positions, and a particle-spin interaction $H_F$.

\noindent
A rigorous characterization of the spin-dependent point interaction Hamiltonian we are going to use in the following is given in  (\cite{ccf}, Theorem 1). 
Here we shall limit ourselves to recall  the statement of the theorem in the one-dimensional case. 

\noindent
Let  $\underline{\alpha}:= (\alpha_{1}, \ldots, \alpha_{N})$ with $ \,\, \alpha_{j} \in \RR^{+} \,\,\forall j \in \mathcal{J} $ be any multi-index of non-negative real numbers; let $\beta, \rho$ be two non-negative real numbers, and let us define  $\underline{\alpha}\cdot \underline{\sigma} := \sum_{j=1}^{N} \alpha_{j} \sigma_{j}$.

\noindent
The operator $H$  with domain 
\begin{multline}\label{fs}
D(H)=\left\{\Psi=\sum_{\underline{\sigma}\in \mathcal S}\psi_{\underline{\sigma}}\otimes\chi_{\underline{\sigma}}\in\mathcal{H}\mid\psi_{\underline{\sigma}}\in H^2(\mathbb{R}\backslash Y)\,\,\,\,\forall\underline{\sigma} \in {\mathcal S};\right. \\\left.
\ds \psi_{\underline{\sigma}}(y_j^+)=\psi_{\underline{\sigma}}(y_j^-)=\psi_{\underline{\sigma}}(y_j)\,, \quad \forall \underline{\sigma} \in {\mathcal S}\,, \,\,\, \forall j \in {\mathcal J} \right.\\\left.
\ds \psi'_{\underline{\sigma}}(y_j^+)-\psi'_{\underline{\sigma}}(y_j^-)= \beta \psi_{\underline{\sigma}}(y_j) - 2\, {\bf i} \sigma_j\,\rho§ \psi_{\underline{\sigma'}}(y_j)\,, \quad \forall \underline{\sigma},{\underline{\sigma'}} \in {\mathcal S}: \sigma_j \neq \sigma_j' ,\,\,\sigma_k = \sigma_k' \,\,\, \forall k\neq j\,
\right\}
\end{multline}
and action 

\be\label{fs1}
H\Psi:=\sum_{\underline{\sigma}\in \mathcal S}\left(-\frac{\hbar^2}{2\,m}\triangle+\underline{\alpha}\cdot \underline{\sigma}\right)\psi_{\underline{\sigma}}\otimes\chi_{\underline{\sigma}}\qquad x\in\mathbb{R}\backslash Y,
\ee
 is a selfadjoint operator.

\noindent
In (\ref{fs}) we made use of the standard notation to specify the limit from the left or from the right, {\it i.e.} $\lim_{y \rightarrow y_j^\pm} \psi_{\underline{\sigma}} (y)=: \psi_{\underline{\sigma}}(y_j^\pm)$.

Few relevant properties of the dynamics generated by the Hamiltonian (\ref{fs})-(\ref{fs1}) are worthy of remark:
\begin{itemize}
\item The evolution of the particle wave packet  is free outside the points where the spins are located;
\item
The quantity $2 \alpha_{j}$ represents the difference in energy between two spin configurations differing only in the value of the $j-th$ spin. For simplicity reasons we will use in the following $\alpha_{j} = \alpha \,\,\,\forall j \in \mathcal{J}$ with $\alpha \in \RR^+$.
\item The parameter $\beta$ represents the strength of the point interaction in any point $y_{j} \in Y$.
\item
The parameter $\rho$ is a measure of the interaction allowing an exchange of energy between particle and spins.
\item The interaction Hamiltonian has non vanishing matrix elements only between states whose spin configurations are equal or differ in one point only. This implies that, at first order in perturbation theory, only transitions of this kind have non zero probability;
\end{itemize}
\vspace{0.3cm}
\noindent
Finally, let us precise the initial condition of the Cauchy problem \eqref{SCH}. We assume that the $\alpha$-particle and the environment are initially decoupled, {\it i.e.} 
\be \label{fsc}
\Psi(0)=(\psi_{{\underline \sigma}_1},0,\cdots,0)^t\,, \quad \psi_{{\underline \sigma}_1}=\psi_{--...--}\,,
\ee
with the quantum particle initial wave packet given by
\be \label{IC}
\psi_{--...--}(0,x):=c [f(x)e^{-{\bf i} {p_0 \over \hbar} x}+f(x)e^{{\bf i} {p_0 \over \hbar} x}]\,,
\ee
 where $c>0$ is a normalization constant,  $p_0$  is the particle average momentum and 

$$
f(x):=
\left\{
\begin{array}{cc}
e^{-{|x|^2 \over 4 \sigma^2}}\,, & x \in (-a,a)\,, \,\,\, a, \sigma \in (0,\infty)\,,\\[3mm]
0\,, & \textrm{elsewhere}\,.
\end{array}
\right.
$$
The initial condition (\ref{fsc})-(\ref{IC}) expresses the fact that at time zero all spins are in the ``down'' state while the particle state is the superposition of two identical gaussian wave-packets moving in opposite directions with average momentum $\pm p_0$. Because of the presence of the Hamiltonian $H_I$ this initial fully decoupled condition state will evolve into an entangled state which cannot be any longer written in product form. \\

Note that the initial condition belongs to the operator domain $D(H)$ (more precisely,  it differs slightly from functions in the domain, due to the truncation of the gaussian). As a consequence, the state of the quantum system evolves remaining constantly in $D(H)$. In particular, the boundary conditions in (\ref{fs}) are satisfied at any time.\\

\section{Numerical discretization} \label{sec2}

Let us now present in this section the numerical scheme we used in order to simulate the evolution of the system described in Section \ref{sec1}, in other words to resolve the Schr\"odinger equation (\ref{SCH}) associated with \eqref{fs}-\eqref{IC}. The results obtained with this scheme will be presented and analyzed in Section \ref{sec3}.\\

At this point, we would like to underline the difficulties in simulating the decoherence phenomenon. The first challenge comes from the limited numerical resources (memory) available in order to take into account for a multi-body quantum environment, in particular more than $N=14$ spin-detectors becomes prohibitively expensive, with the scheme we shall present. To deal with the physically interesting case $N \rightarrow \infty$, one has to think of a different manner of modelling the Wilson-chamber, or to work out an analytical scheme to investigate the asymptotic dynamics of the system as $N \rightarrow \infty$. We plan to come back to this subject in future work.\\
Moreover, the study of the decoherence process relies strongly on the specific system-environment interaction mechanism. In fact, the physical parameters of our model have to be chosen with care, in order to be able to estimate numerically the dynamical evolution of decoherence. In the same sense, even the discretization parameters ($\Delta t,\, \Delta x $) have to be chosen carefully: on one hand, large enough to have tractable numerical simulations, and on the other hand small enough  to be sure to get correct physical results. The choice of all these parameters will be discussed in Section \ref{sec3}.

\subsection{Space-time discretization}  \label{SEC21}
For numerical simulations, we had to truncate the space domain from $\RR$ to $\Omega:=(-L,L)$, $L>0$, and impose boundary conditions in $x=\pm L$. For simplicity reasons homogeneous Neumann boundary conditions are chosen in the following and the simulation domain as well as the simulation time are set such that the $\alpha$-particle is not reaching the border, in order to avoid reflection effects coming from the boundaries. In this manner, one can think of the $\alpha$-particle as evolving on the whole $\RR$-line.\\

Let us now discretize our simulation domain $[0,T] \times \Omega$ in the following homogeneous manner
$$
-L =:x_1 \le \cdots \le x_i \le \cdots x_{N_x}:=L\,, \quad x_i:= -L +(i-1) \Delta x\,, \quad \Delta x := {2L \over N_x-1}\,,
$$
$$
0 =:t_0 \le \cdots \le t_k \le \cdots t_{K}:=T\,, \quad t_k:= k\, \Delta t\,, \quad \Delta t := T/K\,. 
$$
Starting from the known initial condition $\Psi(0,\cdot):[-L,L] \rightarrow \CC^M$, where $M:=2^N$ is the number of possible spin configurations, we are searching at each time step $t_k$, $k=1, \cdots, K$, for an approximation of the vectorial unknown $\Psi(t_k,x_i) \in \CC^M$ in each point $x_i \in [-L,L]$, $i=1, \cdots,N_x$, meaning $N_x * M$ scalar unknowns. This approximation shall be denoted simply by $\Psi_i^k \in \CC^M$.\\

For the points far away from the detectors one can discretize the Schr\"odinger equation, associated with the Hamiltonian given by \eqref{fs1}, via the second-order, unconditionally stable Crank-Nicolson scheme
\begin{equation} \label{NR1}
\begin{array}{l}
\ds {\bf i } \, \hbar \, {\Psi_i^{k+1} - \Psi_i^k \over \Delta t} = - {\hbar^2 \over 4 m } \left( { \Psi^{k+1}_{i+1} - 2\Psi^{k+1}_i + \Psi^{k+1}_{i-1} \over (\Delta x)^2}+ { \Psi^{k}_{i+1} - 2\Psi^{k}_i + \Psi^{k}_{i-1} \over (\Delta x)^2} \right) \\[4mm]
\ds \hspace{7.0cm} + {\mathbb D}_\alpha\, { \Psi^{k+1}_{i} +  \Psi^{k}_{i} \over 2}\,,
\end{array}
\end{equation}
where $i=1,\cdots, N_x$ such that $x_i \notin Y$. Here ${\mathbb D}_\alpha \in \RR^{M \times M}$ is a diagonal matrix whose entries correspond to the different energy levels of the spin-channels and are given by
$$
{\mathbb D}_{\alpha,\underline{\sigma}}:=\alpha \, \sum_{j=1}^N \sigma_j\,, \quad \forall \underline{\sigma} \in {\mathcal S}\,.
$$
Remark moreover that the homogeneous Neumann boundary conditions impose for the ghost points $\Psi^{k}_0=\Psi^{k}_2$ as well as $\Psi^{k}_{N_x+1}=\Psi^{k}_{N_x-1}$ for all $k \in \{ 1, \cdots, K\}$.
This discretization yields  $N_x-N$ vector-equations for the wave-function $\Psi(t_k,x_i) \in \CC^M$ with $x_i \notin Y$.

Missing are now $N$ vector equations.\\

In the points where the detectors are located, i.e. $x_i=y_j$, one has to take into account the effects of the point interaction as well as of the possibility of a crossing to the different channel corresponding to the flipped spin  (see \eqref{fs}). In the following, we shall denote by $i_j\in \{ 1, \cdots, N_x\}$ the index  of the detector $y_j \in Y$, {\it i.e.} $y_j=x_{i_j}$ for all $j \in {\mathcal J}$.
To discretize the particle-environment dynamics in a detector-point $y_j \in Y$, let  us integrate the free Schr\"odinger equation in the two intervals around this point $ (y_j-\Delta x /2,y_j)$ and $ (y_j,y_j+\Delta x/2)$ and sum up the results. This leads to
$$
\begin{array}{lll}
\ds {\bf i}\, \hbar \,  \partial_t \int_{y_j-\Delta x /2}^{y_j} \Psi \, dx &\approx&\ds (\Delta x/2 ) \, {\bf i}\, \hbar\,  \partial_t \Psi(y_j) \\[3mm]
&\approx&\ds - { \hbar^2 \over 2 m} \left[   \partial_x \Psi(y_j^-) - \partial_x \Psi (y_j-\Delta x /2)\right] + (\Delta x /2) \, {\mathbb D}_{\alpha}\,\Psi(y_j) \,,
\end{array}
$$
$$
\begin{array}{lll}
\ds {\bf i} \,  \hbar \,\partial_t \int_{y_j}^{y_j+\Delta x /2} \Psi \, dx &\approx&\ds (\Delta x/2 ) \, {\bf i}\,  \hbar \, \partial_t \Psi(y_j)\\[3mm]
&\approx& \ds - { \hbar^2 \over 2 m} \left[  \partial_x \Psi(y_j+\Delta x /2) - \partial_x \Psi (y_j^+) \right] + (\Delta x /2) \, {\mathbb D}_{\alpha}\,\Psi(y_j) \,.
\end{array}
$$
Summing up these formulae, yields for all $j \in {\mathcal J}$
$$
\begin{array}{lll}
\ds (\Delta x ) \, {\bf i}\,  \hbar \, \partial_t \Psi_{i_j} &= & \ds - { \hbar^2 \over 2 m}\left[ \partial_x \Psi(y_j+\Delta x /2) -\partial_x \Psi (y_j-\Delta x /2) - \partial_x \Psi (y_j^+) +  \partial_x \Psi(y_j^-)\right]\\[3mm]
&& \ds + (\Delta x)\, {\mathbb D}_{\alpha}\,\Psi_{i_j}\,.
\end{array}
$$
Using now the boundary conditions on the detector positions, we get for each $j \in {\mathcal J}$ and each spin-configuration pair $\underline{\sigma},\underline{\sigma'} \in {\mathcal S}$ verifying $\sigma_j \neq \sigma_j'$, and $\sigma_k = \sigma_k'$ for all $k\neq j$,
\begin{equation}\label{NR2}
\begin{array}{lll}
\ds  \, {\bf i}\,  \hbar \, \partial_t \psi_{\underline{\sigma},i_j} &= &\ds - { \hbar^2 \over 2 m}\left[  {\psi_{\underline{\sigma},i_j+1} - \psi_{\underline{\sigma},i_j} \over (\Delta x)^2}-{\psi_{\underline{\sigma},i_j} - \psi_{\underline{\sigma},i_j-1} \over (\Delta x)^2}  -{\beta \over \Delta x} \psi_{\underline{\sigma},i_j}+ {\bf i} {\sigma_j \rho \over \Delta x} \psi_{\underline{\sigma}',i_j} \right]\\[3mm]
&&\ds +  {\mathbb D}_{\alpha,\underline{\sigma}}\,\psi_{\underline{\sigma},i_j}\,,
\end{array}
\end{equation}
yielding after the semi-discretization in time (Crank-Nicolson) the missing $N$ vector-equations.
Let us remark here that the discretization (\ref{NR2}) is very similar to (\ref{NR1}), in particular for $\beta=0$ and $\rho=0$ one gets exactly the free evolution discretization (\ref{NR1}), which is somehow consistent. The terms related to $\beta, \rho$ and $\alpha$ express the fact that there is an energy exchange between the different spin-channels. Note furthermore that the Crank-Nicolson scheme has the essential property of preserving the discrete $||\cdot||_2$ norm, which is a considerable advantage in the present case. \\

The discretization (\ref{NR1})-(\ref{NR2}) gives rise to a sparse matrix, consisting of $M$ tridiagonal blocks, corresponding to the discretization of the Hamiltonian part $H_0+H_D$, and $N$ values per bloc localized outside the blocs and distributed in a well-defined manner, corresponding to the discretization of the particle-detectors interaction part $H_F$. The resolution of this sparse linear system ($(3\,N_x-2)\, M+N\,M$ non-zero elements) has been performed by means of the direct MUMPS solver (LU-decomposition). In the case one wants to increase the number of environmental spins above $N=12$, more performant solver have to be used, as the iterative Krylov solvers, based on preconditionner techniques.

\section{Numerical results} \label{sec3}

Aim of the present section is to use the previously introduced numerical scheme in order to study the creation of ``tracks'' in our simplified Wilson-chamber model. As mentioned earlier, the meaning of pattern formation in the present model is the following: a track is defined as the ionization (spin flip) of more than one atom/spin on only one side of the initial $\alpha$-particle position $x_0=0$.\\
The parameters used for these simulations are summarized in Table \ref{tab}. The choice of the parameters is related to some physical constraints, corresponding to the specific situation we want to describe. In particular, the spin-detectors are divided into two groups, located around $\pm D$, in a symmetric way with respect to the origin of the spherical wave ($x_0=0$). 

\noindent
The parameters will be chosen to satisfy the following assumptions
$$
\beta   \ll 1/d \,\,\,\, \,\,\,\,\,\,\,\,\,\,\,\,\,\,\,\,\, \,\,\,\,\,\,\,\,\,\,\,\,\,\,\,\,\,\,\,\,\,\, d < \sigma \ll D\,.
$$
A small $\beta$ denotes a very weak energy exchange between particle and environment guaranteeing  the non demolition character of the interaction (in fact in dimension 1 it could be put equal to zero). The wave packet initial variance $\sigma$ is chosen of the same order of magnitude  of each spin cluster size. This means that we are in a situation where the particle is interacting simultaneously with the majority of the spins in each cluster. Moreover, the last inequality implies that the support of the wave packet has negligible intersection with the scatterer arrays until a finite flight-time in which it reaches the spin clusters is elapsed. Till then, the flipping probability is going to be negligible. \\

\noindent

Finally, let us remark that the distances and simulation time have been chosen in such a way that before the final time $t_{*}$ the $\alpha$-particle has moved over all spin-detectors, but it did not reach yet the border of the domain, in such a way that disturbing secondary effects (like reflections)  related to the boundaries are avoided.

\begin{table*}[htbp]
\begin{center} 
\begin{tabular}{|c|c||c|c|}
\hline
$L$&$3/2$&$N_x,\,\,  \Delta x$&$1000,\,\, 3*10^{-3}$\\
\hline
$t_\star$&$0.065$&$N_t,\,\,  \Delta t$&$350,\,\,1.8* 10^{-4}$\\
\hline
$\eps$&$10^{-1}$&$\hbar,\, m$&$\eps\,, 1$\\
\hline
$N$&$4,6,8,10,12$&$y_j$&$\pm D \pm (2k+1)\, d/2\,, \,\,\, k=0,1,2,\cdots $\\
\hline
$D$&$L/3$&$d$&$\eps/N$\\
\hline
$x_0$&$0$ &$p_0$&$4/3\eps$\\
\hline
$\sigma$&$\eps/4$&$\beta$&$\eps^4$\\
\hline
$\rho$&$\eps^{-2}$&$\alpha$&$\eps^4$\\
\hline
\end{tabular}
\end{center}
\caption{{\footnotesize Parameters used in the numerical simulations.}}
        \label{tab}
\end{table*}
\vspace{0.3cm}

\begin{figure}[htbp]
\begin{center}
\includegraphics[width=6cm]{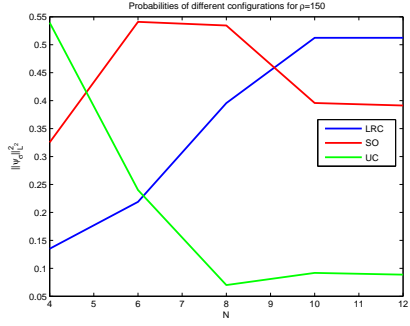}
\end{center}
\vspace{-0.3cm}
\caption{\label{IMA0} 
{\footnotesize Plot of the $2*LRC$, $SO$ and $UC$ probabilities at final time $t_\star$, as a function of $N$ and for $\rho=150$.}}
\end{figure}

\noindent
Starting our simulations with the initial condition given in (\ref{IC}), which corresponds to a situation with all spin-detectors in a ``down''-position, we are firstly interested in the probabilities of some specific spin configurations, at the final time $t_\star=0.065$. The aim is to observe if the configurations corresponding to the creation of a track have a larger probability than other possible configurations. Some of these probabilities obtained with our numerical simulations are summarized in Table \ref{tab1} (for $\rho=100$) and Figure \ref{IMA0} (for $\rho=150$).

{\tiny
\begin{table*}[htbp]
\begin{center} 
\begin{tabular}{c|c|c|c|c}
$\rho=\eps^{-2}$, $N$ & Left/Right Cumulative & One spin & Unchanged  & $2*LRC+OS+UC$  \\\hline 4 &    0.325685025765 E-001   
&    0.275253381822
&    0.659609415084   
& 0.999999802059000   
\\\hline 6 &    0.732817073769 E-001   
&    0.459327397789   
&    0.394108332939  
&   0.999999145481800  
\\\hline 8 &    0.136249083320  
&    0.467653883264   
&    0.259847521850
&  0.999999571754000 
\\\hline 10 
&    0.178222289956
&     0.431768410329
&    0.211787009757
& 0.999999999998000
\\\hline 12 
&    0.211022969661
&     0.429493203402
&    0.148460857272
&    0.999999999996000
\\\hline 14
&0.260042860561 
&0.391267684323
&0.0886465945538
&1.0000000000000
\end{tabular}
\end{center}
\caption{{\footnotesize Sum of probabilities $||\psi_{\underline{\sigma}} (t_\star)||^2_{L^2(\Omega)}$ according to specific configurations $\underline{\sigma} \in {\mathcal S}$, the sum over all configurations $\underline{\sigma} \in {\mathcal S}$ being equal to one.}}
        \label{tab1}
\end{table*}
}
The  quantities listed in Table \ref{tab1} correspond to the following definitions:
\begin{itemize}
\item {\bf N} is the number of spin-detectors distributed symmetrically with respect to the wave-packet initial position ($x_0=0$)
\item {\bf Left/Right Cumulative (LRC)} corresponds to the total probability of all configurations with flipped spins only on one side of $x_0$, excluding the case of a single spin flip, at final time $t_\star$.
\item {\bf One spin (OS)} corresponds to the total probability, at final time $t_*$, of all the configurations, in which only one spin has flipped.
\item {\bf Unchanged (UC)} is the probability of the configuration in which nothing happened during the simulation time, i.e. all the spins remained in the low energy state.
\item  {\bf Multiple tracks (MT)} is the probability to have a spin configuration in which more than one spin on each side moved to a configuration of higher energy $MT=1-2*LRC-SO-UC$.
\end{itemize}
\vspace{0.2cm}
\noindent
What can be observed from the values in Table \ref{tab1} and Figure \ref{IMA0} is that the probability of observing a track on the right or on the left ($2*LRC$) is a strictly increasing function with the number of spins $N$, and that for each $N$ this value is much larger than the probability to have no track formation.
Let us also remark that the sensitivity of the ``device'' is increasing with the number of the detectors, as it is clear from the fact that the probability of having no spin flip ($UC$) is strictly decreasing for increasing $N$.\\

\begin{figure}[htbp]
\begin{center}
\includegraphics[width=6cm]{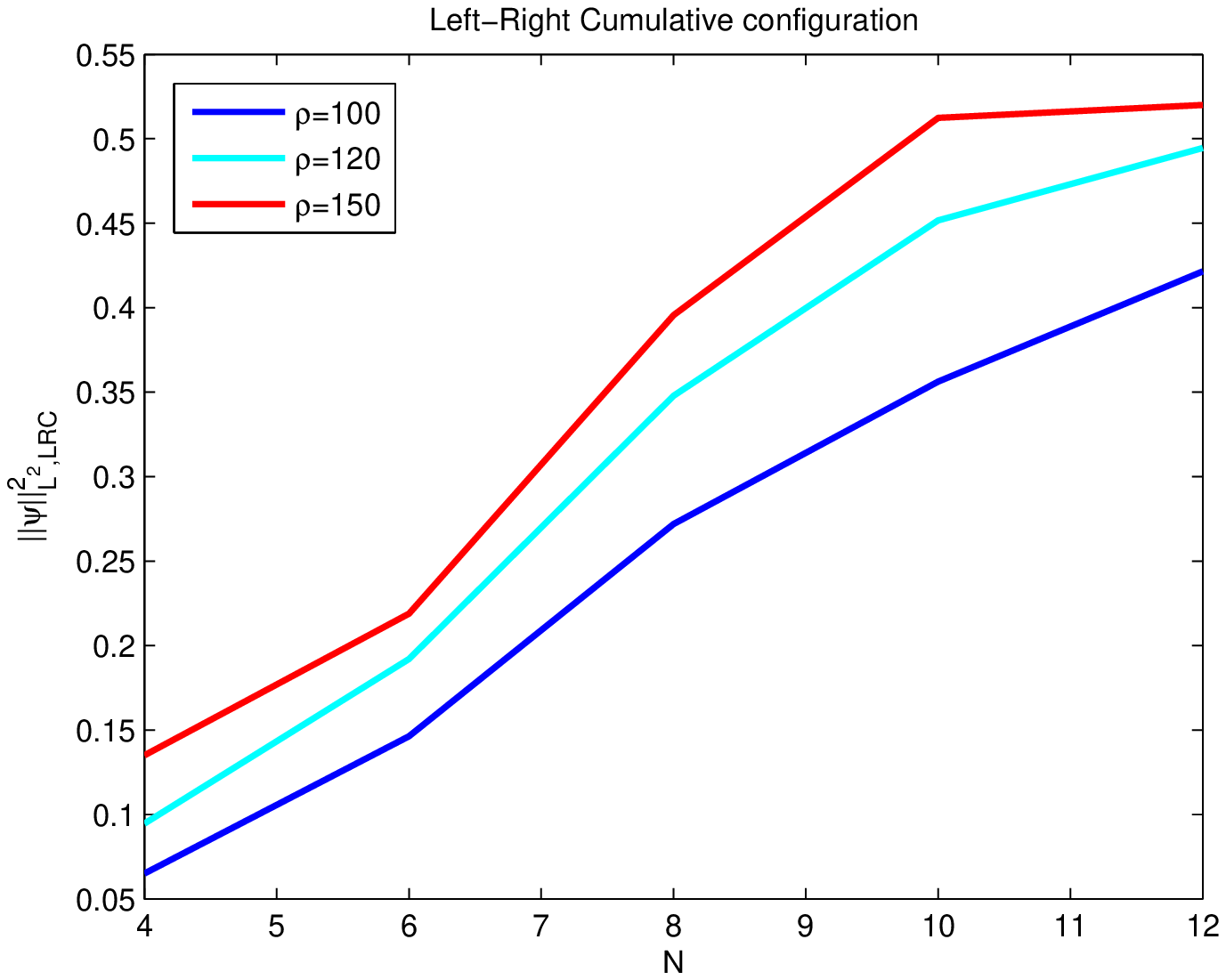}\hfill
\includegraphics[width=6cm]{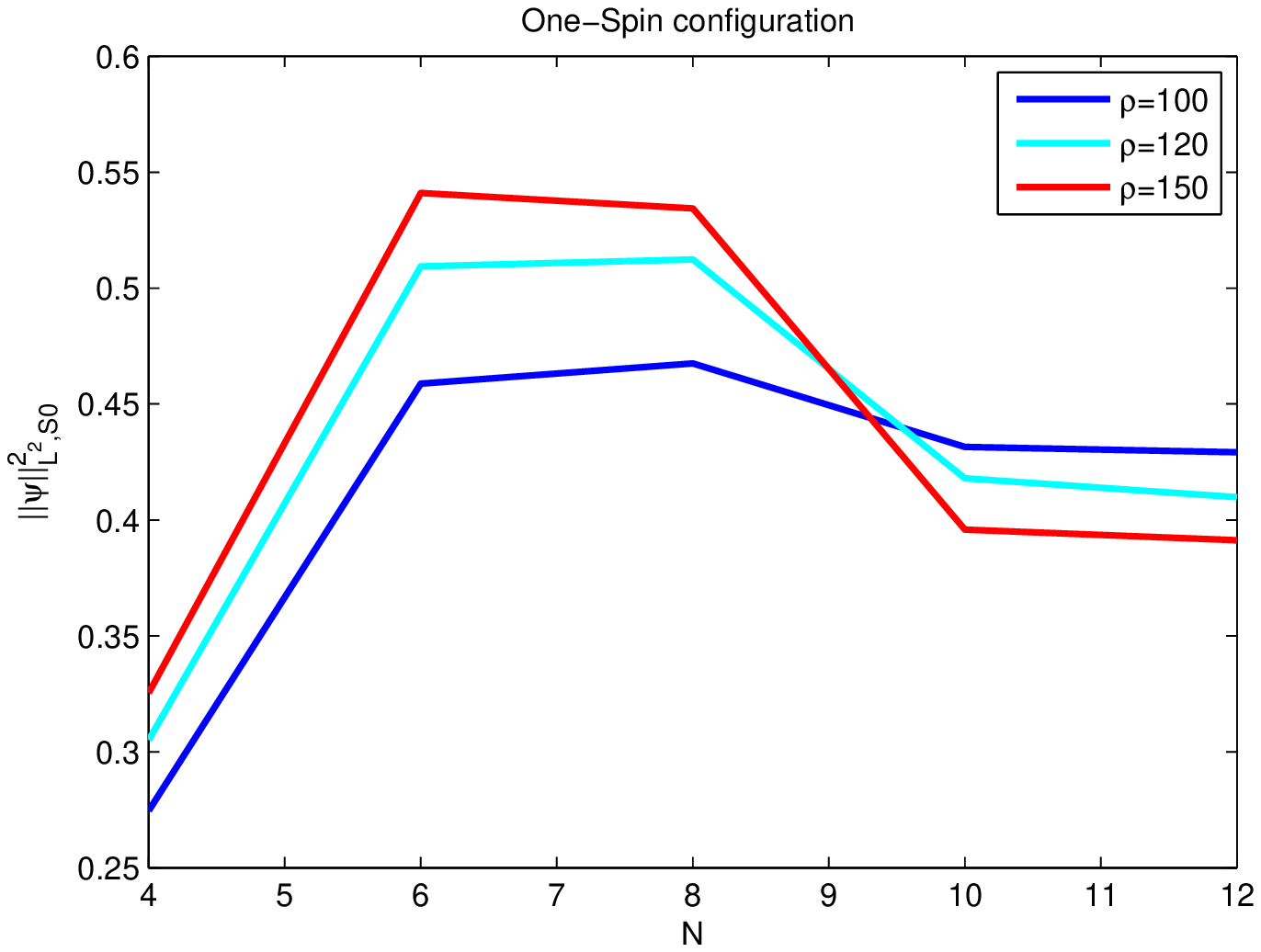}
\end{center}
\vspace{-0.3cm}
\caption{\label{IMA1} 
{\footnotesize Plot of the $2*LRC$ and $SO$ probabilities at final time $t_\star$, as a function of $N$ and for $\rho=100,120,150$.}}
\end{figure}

\begin{figure}[htbp]
\begin{center}
\includegraphics[width=6cm]{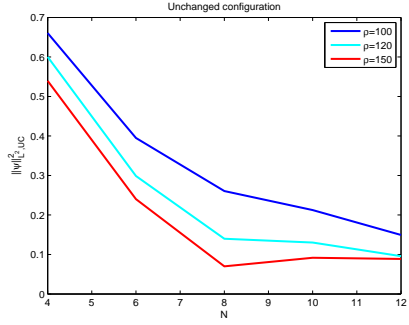}
\end{center}
\vspace{-0.3cm}
\caption{\label{IMA2} 
{\footnotesize Plot of the $UC$ probability at final time $t_\star$, as a function of $N$ and for $\rho=100,120,150$.}}
\end{figure}

\noindent
In order to understand better the influence of the most significant parameters for the decoherence rate enhancement, we carried out other simulations, firstly varying the $\rho$-values for fixed number of spin-detectors $N$. These results are presented in Table \ref{tab_N_R} as well as in Figures \ref{IMA1}-\ref{IMA2}. 

\vspace{0.2cm}
\noindent
\vspace{0.2cm}
\noindent
{\tiny
\begin{table*}[htbp]
\begin{center} 

\begin{tabular}{c|c|c|c|c}
$N=6$, $\rho$ & Left/Right Cumulative & One spin & Unchanged &$2*LRC+OS+UC$  \\
\hline 
150 & 0.109622994819   
& 0.541164020727 
& 0.239589989614 
& 0.999999999979000
\\
\hline
100 &   0.732817073769 E-001   
&    0.459327397789   
&    0.394108332939
&  0.999999145481800
\\
\hline
50 
&    0.103574748581 E-001   
& 0.193411720705
&    0.785873329578 
& 0.999999999999200
\\
\hline
\hline
$N=8$, $\rho$ & Left/Right Cumulative & One spin & Unchanged &$2*LRC+OS+UC$ \\
\hline 
150 &  0.197957193495   
& 0.534889312125 
& 0.691963008808 E-001
& 0.999999999995800
\\
\hline
100 &    0.136249397960
&   0.467653682229     
&    0.259847521850 
&  0.999999999999000
\\
\hline
50 
&    0.200483456577 E-001
&   0.234013722944    
&    0.725889585739 
& 0.999999999998400
\\
\hline
 10 
&    0.433814016219 E-004   
& 0.124587374242 E-001
&   0.987454499772 
& 0.999956618597822
\\
\hline
\hline
$N=10$, $\rho$ & Left/Right Cumulative & One spin & Unchanged &$2*LRC+OS+UC$ \\
\hline 
150 &    0.256357991502   
& 0.396021737286 
& 0.912622797074 E-001
& 0.999999999997400
\\
\hline
100 
&    0.178222289956
&     0.431768410329
&    0.211787009757
& 0.999999999998000  
\\
\hline
50 
&     0.315466612839 E-001   
& 0.268722187965 
& 0.668184489466 
&   0.999999999998800
\\
\hline
\hline
$N=12$, $\rho$ & Left/Right Cumulative & One spin & Unchanged &$2*LRC+OS+UC$ \\
\hline 
150 &  0.260042860561
&  0.391267684323
&   0.886465945538 E-001     
& 0.999999999998800
\\
\hline 
120 &   0.247268176496 
& 0.409755032396
& 0.957086146110 E-001 
& 0.999999999999000
\\
\hline
100 &    0.210785027601
& 0.429140888943
&  0.149289055854
&    0.999999999999000

\end{tabular}
\end{center}
\caption{{\footnotesize Sum of probabilities $||\psi_{\underline{\sigma}} (t_\star)||^2_{L^2(\Omega)}$ according to specific configurations ${\underline{\sigma}} \in {\mathcal S}$, for several $\rho$ and $N$ values. The sum of all configurations (for fixed $\rho$ and $N$) is equal to one.}}
        \label{tab_N_R}
\end{table*}
}

\begin{figure}[htbp]
\begin{center}
\includegraphics[width=6cm]{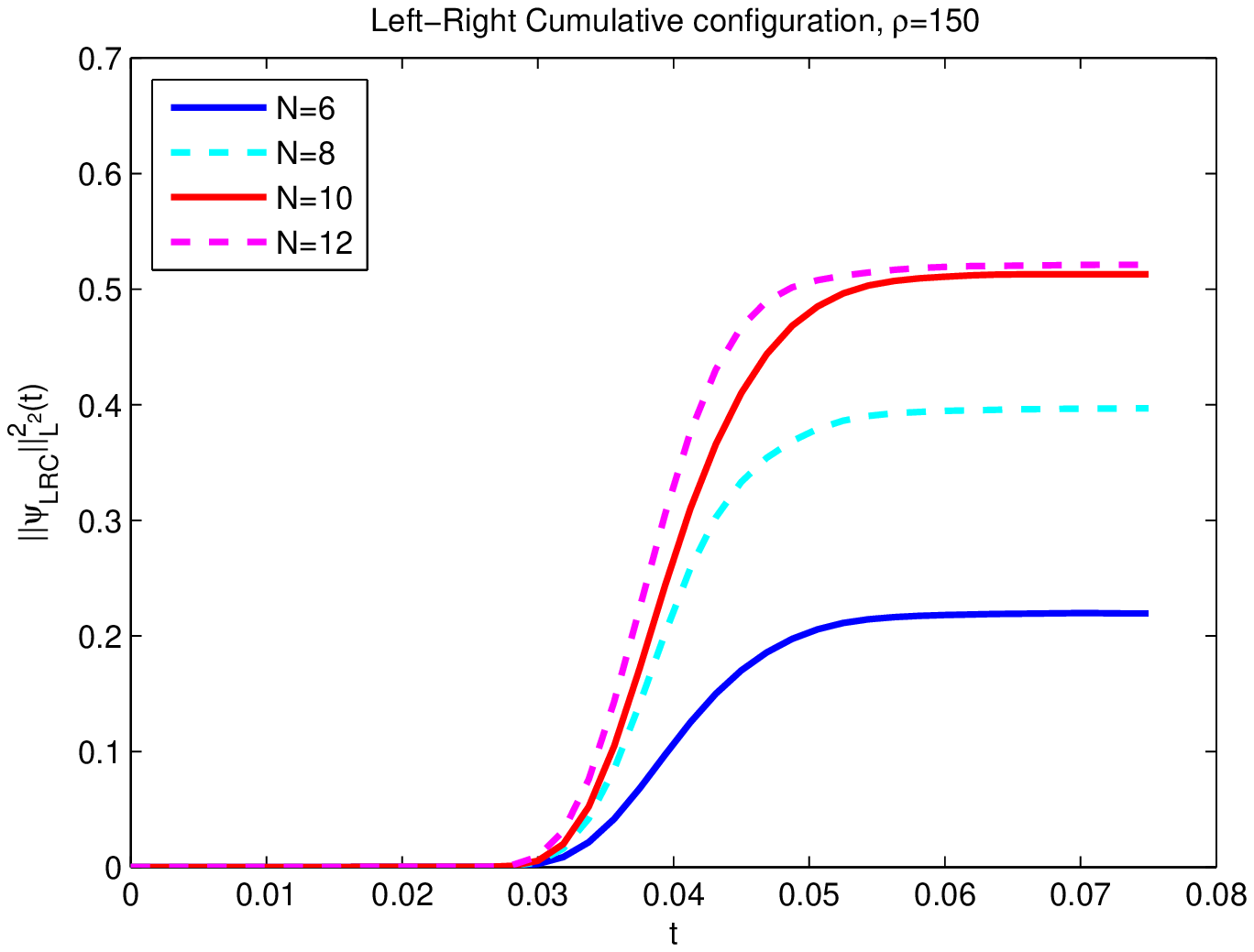}\hfill
\includegraphics[width=6cm]{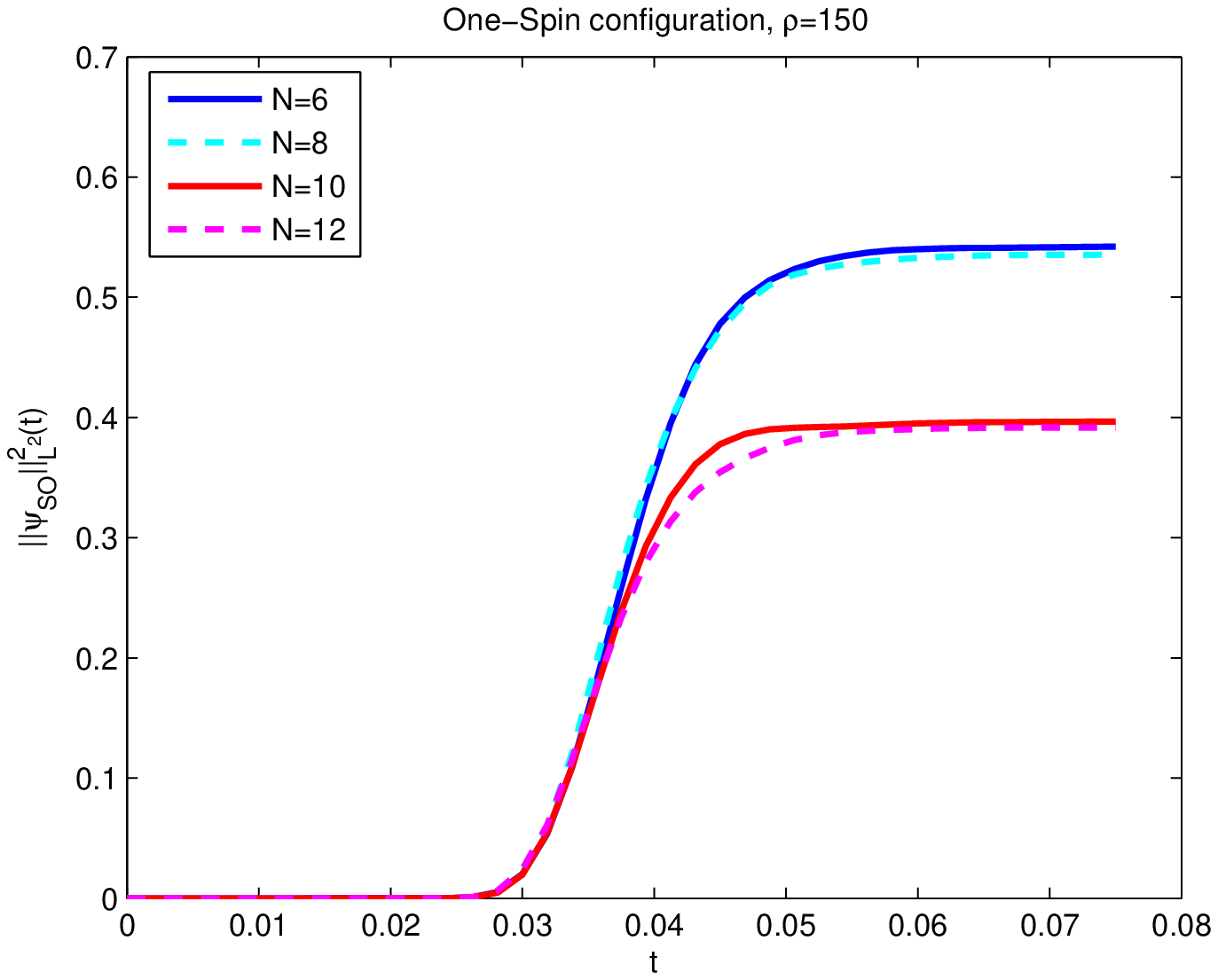}
\end{center}
\vspace{-0.3cm}
\caption{\label{IMA3} 
{\footnotesize Plot of the $2*LRC$ and $SO$ probabilities as a function of $t$, for $\rho=150$ and several spin-detectors $N$.}}
\end{figure}

\begin{figure}[htbp]
\begin{center}
\includegraphics[width=6cm]{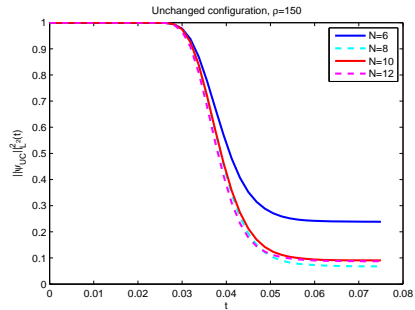}
\end{center}
\vspace{-0.3cm}
\caption{\label{IMA4} 
{\footnotesize Plot of the $UC$ probability as a function of $t$, for $\rho=150$ and several spin-detectors $N$.}}
\end{figure}

In a final study, we were interested in the time-evolution of some  wave-function  components $\psi_{\underline{\sigma}}$ corresponding to a specific spin-configuration $\underline{\sigma} \in {\mathcal S}$. In particular one is interested in the comparison of the probabilities with which the initial state $\psi_{\underline{\sigma}_1}$ with all spins in the low energy state ``down'' ($\underline{\sigma}_1:=\{-\}^N$), transforms into some specific configurations, as for example the Left/Right Cumulative configurations or the One-Spin configurations. 
We represented in Figures \ref{IMA3}-\ref{IMA4} the time evolution of these probabilities $\int_{-L}^L |\psi_{\underline{\sigma}}(t,x)|^2\, dx$, for resp. $N=6,8,10,12$ spin-detectors and $\rho=150$. As expected, the probabilities start to increase at the moment, where the $\alpha$-particle reached the spin-detectors, {\i.e.} at approximately $t=D/p_0=0.0375$.

\vspace{0.2cm}
\noindent
\vspace{0.2cm}
\noindent

\section{Conclusions} \label{sec4}

Our aim was to give numerical results concerning the evolution of a quantum particle in a quantum environment. We modelled the environment as an array of $N$ localized two level quantum systems interacting with the particle, as soon as the particle wave  function is different from zero on their (fixed) positions. We considered a spatially one dimensional model where the $N$ constituents of the environment are  distributed symmetrically with respect to the origin: $N/2$ of them on the right side of the origin and $N/2$ of them localized in symmetric positions on the the other side of the the origin. The interaction hamiltonian was chosen in the family of the so called multi-channel point interaction hamiltonians, extensively used, since decades, in applied quantum physics (\cite{lo},\cite{do},\cite{sepvs}\cite{ccf},\cite{ccf3},\cite{ft}).

\noindent
The evolution of the whole system, made of the particle and of the spins, is completely represented as a multi-channel wave function for the particle, each channel corresponding to one of the $2^{N}$ possible quantum states of the array of spins. The initial state is chosen to belong to the channel where all the spins are in the down state (the one with minimal energy). The interaction hamiltonian allows crossing of channels when the particle has non zero probability to be in the region occupied by the spins.

\noindent
The numerical analysis we performed shows that, as expected, the state of the whole system, after a short interaction time and apart from negligible terms, is the sum of three wave packets corresponding to three ``macroscopically'' different states: 
\begin{enumerate}
\item no significative change in the overall spin state of the array has taken place and the particle is approximately in its initial state,
\item a  ``substantial percentage'' of spins on the right side of the origin changed state and the particle is going to the right,
\item a  ``substantial percentage'' of spins on the left side of the origin changed state and the particle is going to the left.
\end{enumerate}

We computed independently the wave function in the channel where only one spin flipped with respect of the initial condition. On one hand, one point is always on a single side with respect of the origin, on the other hand, taken into account the total number of spins we are able to manage, one is not a small percentage of points on one side of the origin. Nevertheless, we chose to consider the flipping of only one spin as part of case $(1)$ (nothing happened to the environment). 

\noindent
In any case, our analysis suggests that the probability of $(1)$ is decreasing when the number of constituents of the environment increases. In terms of the environment evolution, the numerical results indicate that the only evolutions of the environment with non negligible probabilities are 
\begin{itemize}
\item no ``tracks'', 
\item a ``track'' on the left of the origin, 
\item a ``track'' on the right of the origin.
\end{itemize}

\noindent
In fact, what our results show is that the probability of multiple spin flips on both sides of the origin is negligible at all times. 

\noindent
We plan to examine in further work some technical and fundamental open problems in the evolution of a quantum particle in a quantum environment. First of all, we would like to push the computation to the point where $N$ is sufficiently large to allow to specify rigorously the meaning of ``macroscopic  change of the environment '' with respect to the initial conditions. We then would like to face the problem of analyzing the dependence of the results on the basis we use to represent the initial state of the particle.

\noindent
Last, but not least, we plan to analyze the problem in dimension larger than one. For $d=2$ and $d=3$ a general definition of multi-channel point interaction is available . The numerical analysis in those cases is probably simplified considering the dynamical equations governing the singularities of the wave functions on the points where the spins are localized (for details see e.g \cite{martina}).

\vspace{0.3cm}

\noindent {\bf Acknowledgments.} The authors would like to acknowledge
support from the ANR LODIQUAS  (Modeling and Numerical Simulation of
Low Dimensional Quantum Systems, 2011-2014) and FIR grant ÒCond-
MathÓ 
RBFR13WAET.

\end{document}